# ProtoCode: Leveraging Large Language Models for Automated Generation of Machine-Readable Protocols from Scientific Publications

Shuo Jiang[1,ø], Daniel Evans-Yamamoto[2,ø], Dennis Bersenev[1], Sucheendra K. Palaniappan[2,*] and Ayako Yachie-Kinoshita[1,2,*]

1. SBX BioSciences, Inc., 1600 - 925 West Georgia Street, Vancouver, BC, V6C 3L2, Canada

2. The Systems Biology Institute, Saisei Ikedayama Bldg., 5-10-25, Higashi Gotanda, Shinagawa-ku, Tokyo, 141-0022, Japan

ø These authors contributed equally.

* Correspondence should be addressed to SKP (sucheendra@sbi.jp) and AYK (yachie@sbx-biosci.com).

## Abstract

Protocol standardization and sharing are crucial for reproducibility in life sciences. In spite of numerous efforts for standardized protocol description, adherence to these standards in literature remains largely inconsistent. Curation of protocols are especially challenging due to the labor intensive process, requiring expert domain knowledge of each experimental procedure. Recent advancements in Large Language Models (LLMs) offer a promising solution to interpret and curate knowledge from complex scientific literature. In this work, we develop ProtoCode, a tool leveraging fine-tune LLMs to curate protocols which can be interpretable by both human and machine interfaces. Our proof-of-concept, focused on polymerase chain reaction (PCR) protocols, retrieves information from PCR protocols at an accuracy ranging 69-100% depending on the information content. In all the tested protocols, we demonstrate that ProtoCode successfully converts literature-based protocols into correct operational files for multiple thermal cycler systems. In conclusion, ProtoCode can alleviate labor intensive curation and standardization of life science protocols to enhance research reproducibility by providing a reliable, automated means to process and standardize protocols. ProtoCode is freely available as a web server at https://curation.taxila.io/ProtoCode/.

# 1. Introduction

The field of life science research is continuing to be developed in recent years, with over 36 million citations in the PubMedalone (1). Typically, publications use and/or adapt data, biological resources, and methods from previous literature. Adherence to the Findable, Accessible, Interoperable, and Reusable (FAIR) principles for data and experimental protocol details is crucial for enhancing research (2). With the increasing amount of data and literature, providing information in forms which machines can automatically find and use is an important aspect for resources. Databases such as The Sequence Read Archive (3) and Protein Data Bank (4) exemplify this trend, requiring specific data formats and metadata to enhance machine-based reusability. A major biological resource distributor, Addgene (https://www.addgene.org/), now encourages depositors to upload files with data describing the source and construction process of deposited synthetic DNA constructs (5), in addition to the traditional metadata. While the accessibility for research data, analysis scripts, and protocols have become common practices, the reusability of protocols remains mostly limited to human use. The extension of automated machine reusability on protocols would open the gate to exploit the rich literature for applications such as lab automation.

Protocols are typically described in the form of text in scientific publications. Some peer-reviewed journals require the authors to describe the resources and methods in a standardized format to ensure that all necessary information is available to reproduce the protocol (6). Journals dedicated to protocols and repositories such as protocol.io (https://www.protocols.io/) provide a platform to describe protocols in detail with additional notes. Videos are also used to visually aid researchers to replicate specific techniques (https://www.jove.com/). While these standardization and guidelines increase reproducibility and reliability of the described protocol, there remain some limitations. The first limitation is that the current models/guidelines have large variability that even the same procedures in two protocols can be described differently. The second limitation is that the protocols described using any of the guidelines above cover only a small fraction of present literature, since only new publications in specific journals follow such guidelines. An approach to circumvent these limitations is to curate existing protocols from the current literature, generating standardized protocols in machine readable formats. Previous efforts have proposed programming languages and data formats to describe protocols with no ambiguity (7) (https://autoprotocol.org/, https://github.com/Bioprotocols/labop). Ideally, curation and conversion of protocols in the literature to such machine readable formats would benefit the research community, but is challenging due to a labor intensive process requiring specialists with domain expertise to perform such tasks.

In recent years, Large Language Models (LLMs) have seen remarkable advancements, demonstrating their ability to extract, comprehend, curate and summarize human-written natural language, substantially reducing the time spent by researchers on repetitive tasks. Specifically, in the life science field, LLMs have begun to demonstrate significant impact by encoding clinical knowledge for medical applications (8) or identifying biological functions of gene groups (9, 10). LLMs have also been applied to design and evaluate research such as automatic evaluation of long-term experiment plans (11), creating robot operating scripts for lab automation (12, 13), and *de novo* PCR primer design (14). In the perspective of protocol sharing, LLM assistance has been applied to assist script generation to run on cross-platform frameworks which are compatible with multiple liquid-handling robot systems (15). The curation and conversion of protocols in the literature to machine readable formats have been attempted in the field of chemistry, where LLM was used for extracting FAIR data on chemical experiments (16). Leveraging LLM to parse FAIR data on life science experiments has been challenging due to the diversity and complexity of parameters between experiments which requires task-specific models for parsing data for different experiments.

In this paper we introduce ProtoCode, a pipeline that leverages LLMs to transform human-written, non-standardized, natural language descriptions of life science protocol to machine readable formats that can serve as intermediate representation formats between natural language and equipment operation files. Using Protocode, we curated 200 papers describing protocols of a basic experimental procedure, the Polymerase Chain Reaction (PCR) (17), and demonstrated efficient curation. The underlying LLM in ProtoCode is a task-specific model, fine-tuned on these 200 expert-curated PCR protocols atop the base GPT-3.5-turbo model, thereby enhancing its capability to accurately interpret and standardize scientific protocols. The output from ProtoCode fine-tuned model is versatile, where users can generate program files to operate PCR thermal cyclers. Taken together, we demonstrate that ProtoCode is a general pipeline to adopt, reuse, and build on the methods described in current literature. ProtoCode is freely available as a web server with no login requirement at https://curation.taxila.io/ProtoCode/.

## 2. Results

**2.1 The ProtoCode pipeline**

We developed ProtoCode, a pipeline dedicated to process life science protocol (in this instance, for PCR) in natural language text. Its two main features are (i) the extraction of essential information of protocol descriptions in natural language text and (ii) format conversion to downstream applications (**Figure 1A**). For each protocol, ProtoCode fine-tunes a LLM for information extraction using curated data (**Figure 1B**). This fine-tuned LLM can parse natural language protocol descriptions using a predefined, protocol-specific intermediate representation format, where all information is stored. The parsed data, in JSON format, is a versatile format where researchers can convert to various formats including, but not restricted to, standardized natural language text, tables, and equipment operation files. The ProtoCode pipeline is implemented as a web server available at https://curation.taxila.io/ProtoCode/ where it is publicly available and free to use (**Figure 1C**).

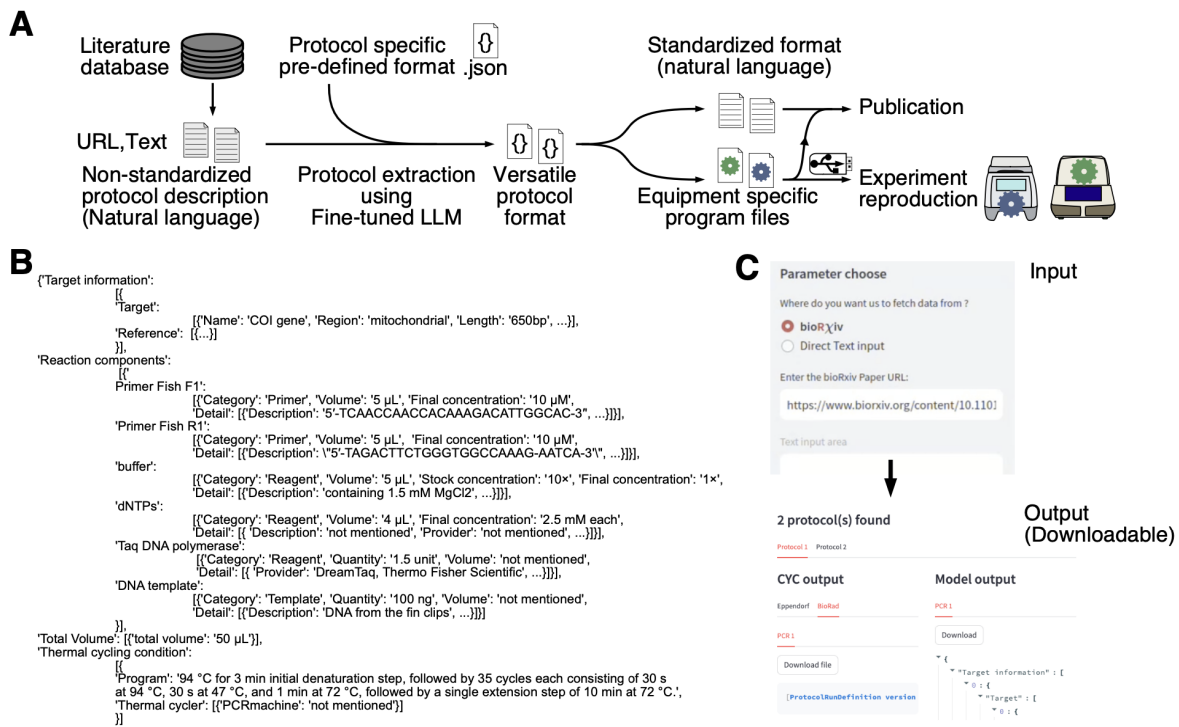

**Figure 1. Schematic overview of the ProtoCode workflow.**

(**A**) Overview of the ProtoCode pipeline. Protocols described in natural language will be parsed using a versatile format which contains data entries specific to each protocol. The parsed protocol in the versatile format can then be converted to natural language for standardized nomenclature, or program execution files for experimental equipment.

(**B**) Example of a PCR protocol described in JSON format. Text data are extracted and formatted into the experiment specific predefined JSON format. Entries which were not contained in this specific example were abbreviated.

(**C**) User interface of the ProtoCode web server. The user can input either a biorxiv URL or text. Resulting ProtoCode outputs are displayed and are available for download.

**2.2 ProtoCode usage**

To begin protocol curation using ProtoCode, users need to submit their URL for the literature described in natural text, and choose the fine-tuned LLM based on the protocol to be extracted using a dropdown menu. Upon submission, ProtoCode first conducts a screening analysis to identify the text region corresponding to the specified protocol. Next, ProtoCode performs data extraction using the protocol specific fine-tuned LLM. The extracted data is in the form of JSON format, with all necessary information. Moreover, if the corresponding protocol data contains information on equipment settings and/or programs, users can select the outputs for operating experimental equipment.

**2.3 ProtoCode successfully retrieves information on PCR protocols**

We performed a proof-of-concept demonstration of the ProtoCode pipeline on a general technique, PCR (17). We designed a data structure for the versatile protocol format to store information of PCR procedures (**FIgure 1B**), and curated 200 non-standardized protocols from the literature database. Using this data, we performed a 5-fold cross validation of PCR specific fine-tuned LLM (**Figure 2A**). The fine-tuned LLM could accurately retrieve information on total reaction volume, PCR program, and thermal cycler programs (accuracy and recall of 100%). On the other hand, for more complex cases of multi-component data entries such as reaction component and amplification target, the performance was somewhat lower, with accuracy of 88.7% and 69.2% and recall of 88.7% and 59.0%, respectively (**Figure 2B**). The accurate extraction of total reaction volume, PCR program, and thermal cycler program information is consistent with the string match ratio, where in all cases identically matched with the manual curated validation dataset (mean ± sd: 100 ± 0 %) (**Figure 2C**). The fine-tuned LLM showed imperfect retrieval of multi-component data for reaction component and amplification target. The intersection over the union values of detected reaction components and target regions were on average 0.83 and 0.60, respectively (**Figure 2D**).

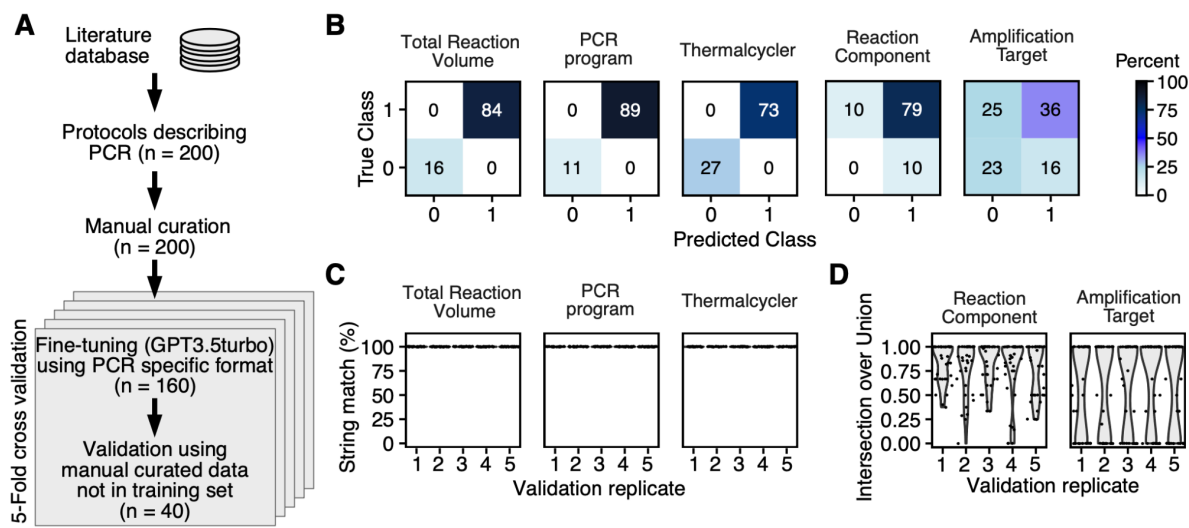

**Figure 2. Benchmarking ProtoCode pipeline on curating PCR protocols.**

(**A**) Validation strategy for ProtoCode, on PCR protocols. A total of 200 protocols were retrieved from literature sources, and were manually curated using the pre–defined format. We employed a 5-fold cross validation for fine-tuning and quality assessment.

(**B**) Confusion matrix for detection of data entries specified in the format for PCR curation. For datasets where multiple components are present (Reaction Component and Amplification Target), values were weighted based on the number of elements so that the total value for each protocol equals one. Refer to **Figure 1B** for data nomenclature. Rounded percent values, which represent the fraction of protocols in each quadrant, are annotated.

(**C**) Violin plot of string match ratio of detected data entries for total PCR volume (Left), PCR program (Center), and thermal cycler (Right).

(**D**) Violin plot of intersection over union values of multi-component entries in each PCR protocol. (Left) Values for reaction components for the PCR. (Right) Values for target regions subjected to PCR amplification.

## 2.4 ProtoCode enhances reproducibility by generating equipment operation files

To maximize the benefit of the retrieved PCR program information from protocols, we developed a component within ProtoCode designed to transform parsed data into operation files for thermal cyclers for downstream use. We implemented this for thermal cyclers from two major brands, BioRAD (California, United States) and Eppendorf (Hamburg, Germany). We validated the success rate and accuracy for generating such PCR program files from ProtoCode outputs using protocol descriptions related to prior publications (18–20), where the PCR program files are accessible for validation (**Figure 3A**). Validation of the ProtoCode output was performed using the pre-existing PCR program files to operate thermal cyclers from BioRAD and Eppendorf brands. We show that in all tested cases, ProtoCode output could be converted to formats to operate each of the two thermal cycler brands, regardless of the thermal cycler used in the original publication (**Figure 3B** and **Supplementary Table 2**). We note that protocols used for this validation were not included in the original dataset for fine-tuning and/or validation. Taken together, we show that our ProtoCode pipeline is capable of enhancing research reproducibility by generating program files to operate equipment.

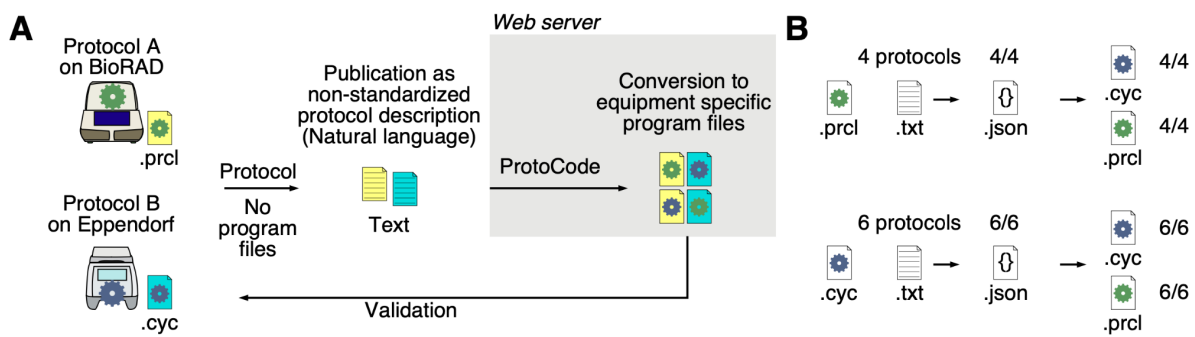

**Figure 3. Versatile format conversion of curated protocols.**

(**A**) Schematic illustration of thermal cycler operation files generated from ProtoCode output. Published text descriptions of PCR programs are extracted using ProtoCode, and converted to equipment specific operation file formats. Validation was performed using pre-existing PCR program files to operate thermal cyclers from BioRAD and Eppendorf brands.

(**B**) Results from the validation for generation of PCR program files from ProtoCode output.

## 3. Discussion

In this study, we present ProtoCode, an LLM-based pipeline capable of extracting protocol information from non-standardized natural language text. ProtoCode enables users to convert extracted information to both human and machine readable formats, serving as intermediate representation formats between natural language and equipment operation files. The usability of ProtoCode was demonstrated on a basic experimental procedure, PCR. The capability of ProtoCode to generate thermal cycler operation files in multiple formats shows its practical application for reproducing experiments. Our demonstration of ProtoCode on PCR shows that further implementation of other protocols can lead to comprehensive standardization of protocol descriptions in life science research. ProtoCode can be applicable to other protocols by creating a protocol specific intermediate representation format and a fine-tuned LLM for content curation. Further works are to be followed for the number of protocols and output formats implemented for ProtoCode curation.

The accurate and complete description of protocols is the basis of reproducible research. In our pipeline, this relies on two factors. The first is the protocol specific intermediate representation format used for LLM fine-tuning. Defining a format to describe specific protocols has been challenging for manual curation due to the nature of incomplete and/or unambiguous description in available protocols. The incompleteness of protocol description requires a specialist with domain expertise to deduct the actual procedures, which can cause large variation between curators. Curation in machine readable formats could circumvent this issue of human dependent variability through data mining, where a complete list of parameters can be identified. The list of parameters can be used to define the template format to describe protocols, where it can serve as a guideline for new descriptions. The data mining strategy on protocols would also alleviate the interpretation and reproduction of protocols with missing information, where the lacking parameters can be inferred based on other properties of the protocol (e.g. sample and reagent properties). We believe our approach can serve as a basis for future optimization of protocol description through community effort. The second is the performance of the protocol specific fine-tuned model to extract all information accurately and completely. Our observations from the fine-tuned LLM on PCR showed varied performance of information extraction between parameters, where multi-component information (reaction components and amplification targets) performed in low accuracy of 69.2-88.7%. This may be a result from the high variability of content and description, which requires larger amounts of data to train the fine-tune LLM. Further investigation is required to determine the amount of data to achieve accurate and complete retrieval of parameters for each protocol.

Finally, we believe intermediate representation of protocols from ProtoCode show potential for application to lab automation setups. Previous studies have demonstrated that LLM can be leveraged for automatic evaluation of long-term experiment plans (11) and robot operating scripts for lab automation (12). The current ProtoCode curation is limited to specific experiments, where conversion of protocols to equipment operation files is only compatible with efforts of general automation settings where the same experimental equipment is used (21, 22). Further investigation is required for applying ProtoCode to curate serieses of experiments, which would allow conversion to multiple automation pipelines and framework languages (23–27).

## 4. Material and Methods

**4.1 Data Collection and Preprocessing**

We collected literature of open access and open archives from ScienceDirect (https://www.sciencedirect.com/) and BioRxiv (https://www.biorxiv.org/), using the provided API. We retrieved literature containing description of reverse transcription-quantitative PCR (RT-qPCR) and general PCR protocols. The 'Material and Methods' sections of papers were evaluated using a custom python script for extracting the text containing information of interest. The text corresponding to PCR protocols were identified by selecting subsections and text regions using regular expressions matches using keywords shown in **Supplementary Note 1**. In total, we retrieved PCR protocols from 200 articles (**Supplementary Table 1**).

### 4.2 Fine-tune model construction

LLM training data was prepared in a JSON format using the GPT-3.5-turbo-0613 model on the OpenAI API using a pre-defined format specific for PCR protocols. The 'system' role for the GPT model was configured as a biomedical expert. See **Supplementary Note 2** for the full prompt description. The GPT output was manually corrected for any errors. The resulting training dataset in JSON format was used to fine-tune the GPT-3.5-turbo-0613 LLM (Open AI, California, United States).

### 4.3 PCR protocol format conversion

Extracted PCR protocol information in JSON format was converted to formats for downstream applications using custom python scripts. The PCR program information of each protocol were converted to protocol files for program execution in thermal cyclers for Eppendorf (Hamburg, Germany) and Bio-Rad Laboratories (California, United States) in .cyc and .prcl extensions, respectively. Regular expression and named entity recognition were used to obtain the temperature, duration time, and total reaction volume from text, using the python package re (version 3.12.0).

### 4.4 Validation of ProtoCode

Validation was performed by comparing manual curated protocols to the fine-tuned LLM output for the same protocol. This was done in a five-fold cross-validation, where one subset (n = 40) was used for validation, while the remaining subsets represent the training data (n = 160). For each of the data categories collected, string similarities were calculated using the WRatio function of the python package fuzzywuzzy (version 0.18.0). Perfect string match was defined values over 95. For multi-component data entries, the intersection over union values were used to compare the agreement between the two sets.

### 4.6 Web server implementation

The web server implementation utilizes Streamlit, a Python library known for its simplicity and interactive capabilities. Running on an Amazon EC2 instance with a static IP, the system deploys Streamlit to create a user-friendly interface. Users can directly input textual inputs or link to bioRxiv URL, which internally triggers the preprocessing, extraction and model inference using the fine tuned model to extract experimental protocols. The backend of our system is fully implemented in Python.

## Data availability

The list of literature to fine-tune the GPT-3.5-turbo-0613 LLM for PCR data extraction is shown in **Supplementary Table 1**. The example dataset is available for download on the tutorial page of our website. Detailed documentation and instructions for using ProtoCode is publicly available at https://github.com/leisuzz/ProtoCode.

## Acknowledgements

The authors thank Dr. Ajay Sethi for insightful comments on the manuscript.

## Declaration of Conflicting Interests

The authors declare no competing interests.

## Funding

This work was supported by the ONRG Grant for the Nobel Turing challenge to The Systems Biology Institute (Grant number: N62909-21-1-2032).

## Author contributions

SJ, SKP, and AYK conceived the study. SJ constructed the LLM and subsequent pipelines with support from DEY and DB. DEY performed the validation analysis. SKP implemented the web server. SKP and AYK supervised the work. DEY and SJ wrote the original draft of the manuscript. DEY, SKP, and AYK edited the manuscript. All authors have approved the final manuscript.